\def\half{\frac{1}{2}}
\documentclass{iopart}\usepackage{cite}\usepackage{graphicx}
\begin{document}
\markboth{R.~L.~Hall \& W.~Lucha}{Semirelativistic stability of
boson stars and tightened bounds on their critical mass}

\title{Semirelativistic stability of $N$-boson systems bound by
$1/r_{ij}$ pair potentials}

\author{Richard~L.~Hall}
\address{Department of Mathematics and Statistics, Concordia
University,\\ 1455 de Maisonneuve Boulevard West, Montr\'eal,
Qu\'ebec, Canada H3G 1M8\\ rhall@mathstat.concordia.ca}

\author{Wolfgang~Lucha}
\address{Institute for High Energy Physics, Austrian Academy
of Sciences,\\ Nikolsdorfergasse 18, A-1050 Vienna, Austria\\
wolfgang.lucha@oeaw.ac.at}

\begin{abstract}We analyse a system of self-gravitating identical
bosons by means of a semirelativistic Hamiltonian comprising the
relativistic kinetic energies of the involved particles and added
(instantaneous) Newtonian gravitational pair potentials. With the
help of an improved lower bound to the bottom of the spectrum of
this Hamiltonian, we are able to enlarge the known region for
relativistic stability for such boson systems against
gravitational collapse and to sharpen the predictions for their
maximum stable mass.\end{abstract}


\vspace{3ex}

\vskip 0.2in \maketitle

\section{Introduction}In this paper we study the implications of
two aspects of relativistic bound systems: the Coulomb (or
gravitational) one-body coupling limit, and the effective coupling
enhancement induced in a system of many identical particles
interacting pairwise. These two effects lead to the conclusion
that a system of $N$ identical particles interacting by attractive
$1/r$ pair potentials becomes unstable if $N$ is very large. Our
principal goal is to sharpen previous bounds on the critical mass
of such a system.

Relativistic quantum-mechanical theories imply an upper limit on
the strength of the coupling of a single particle bound by an
attractive Coulomb potential. Thus for a Hydrogen-like
one-particle system with mass $m$, and units such that $\hbar = c
= 1,$ the upper limits to the allowed coupling $v$ in the
potential $-v/r$ are, respectively, $v < 1$ for the Dirac equation
and $v < \frac{1}{2}$ for the Klein--Gordon equation. Meanwhile,
for the semirelativistic Salpeter equation \cite{BSE,SE,Lieb96}
with Hamiltonian $h = \sqrt{p^2+m^2} -v/r,$ Herbst \cite{Herbst}
showed that for $v < 2/\pi$ the spectrum of $h$ in $[0,m)$ is
discrete and, moreover, he found an explicit lower bound. In
summary
\begin{equation}
h = \sqrt{p^2+m^2} -v/r~ > ~ m\sqrt{1-(\pi v/2)^2},\quad v <
\frac{2}{\pi}. \label{Eq:HLB}
\end{equation}
Under the Schr\"odinger equation, with one or more particles,
there is no such coupling restriction; thus the existence of such
a coupling-limit is essentially a relativistic phenomenon. The
Salpeter Hamiltonian has eigenvalues that lie between the
corresponding Schr\"odinger and Klein--Gordon energies. Thus, in
addition to exhibiting the relativistic coupling limit, within the
allowed couplings, the Salpeter energies are intermediate between
those of Schr\"odinger and Klein--Gordon. For example, for the
one-body problem with mass $m=1$ and the potential $V(r) = -v/r,$
the three theories have ground-state eigenvalues that depend on
the coupling $v$ as shown in Fig.~1. The Salpeter result was
obtained by the use of a scale-optimized trial function with
coordinate expression $\phi(r) = ce^{-r/a};$ it is known
analytically that the exact Salpeter curve is bounded below by the
Klein--Gordon results for $v < \half.$ A brief review of aspects
of Salpeter semirelativistic theory may be found in
Ref.\cite{facets}.
\begin{figure}[htbp]\centering\includegraphics[width=12cm]{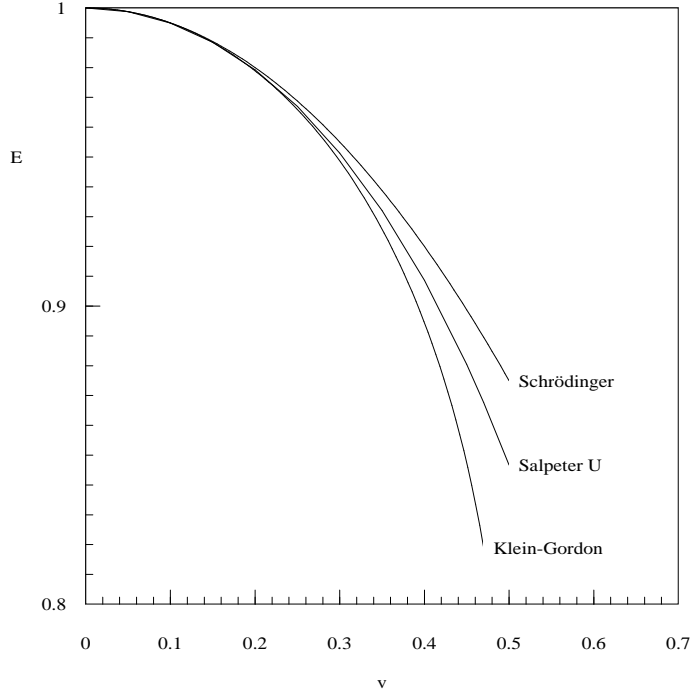}
\caption{Ground-state energies $E$ (in dimensionless units) for
the potential $V(r) = -v/r$ according to the Schr\"odinger,
Salpeter, and Klein--Gordon theories. The Salpeter curve is a
variational upper bound. }\label{Fig:E(v)}\end{figure}
If we consider, as we do in this paper, a system of $N$ identical
particles interacting pairwise via attractive potentials of the
form $-v/r_{ij},$ then the necessary permutation symmetry of the
wave function effectively enhances the pairwise coupling by a
factor of the order of $N.$ This effect, which we shall soon make
clear, is most pronounced in the case of bosons.

Particle identity in quantum mechanics is so strong that, in a
system of identical particles, the particles lose their
individuality; they cannot be separately tracked. This is often
helpful for many-particle theory since, when it comes to
permutation symmetry, at most two of the many possible Young
Tableaux need be considered; moreover, many quantities are
necessarily equal on the average. We shall now make clear the
notion of effective many-body enhancement of the pair couplings
which we alluded to above. We do this in the context of the
problem that is the main concern of the paper. One of the
advantages of the Salpeter semirelativistic theory is that it
accommodates a straightforward formulation of the many-body
problem. We consider therefore a semirelativistic system of $N$
self-gravitating identical bosons of mass $m$ and momenta ${\bf
p}_i,$ $i=1,2,\dots,N.$ This can be described --- in a Newtonian
approximation, justified to some extent by the assumption of a
weak gravitational field --- by the Hamiltonian
\begin{equation}H=\sum_{i=1}^N\sqrt{{\mathbf
p}_i^2+m^2}-\sum_{1=i<j}^N\frac{\kappa}{r_{ij}},\quad\kappa>0,
\label{Eq:HN}\end{equation}
where, in the Newtonian pair potential,
the gravitational interaction strength (determined by the
gravitational constant $G$ and the particle mass $m$) has been
encoded in the coupling parameter $\kappa := Gm^2$. The pair
distance between the interacting particles $i$ and $j$ is given by
 $r_{ij}\equiv|{\mathbf x}_i-{\mathbf
x}_j|.$ If we consider expectations with respect to a normalized
boson function $\Psi$, then we immediately find that there is a
relation between $H$ and a scaled two-particle Hamiltonian,
namely, $\langle H\rangle\ = \langle H_2\rangle$, where
\begin{equation}
H_2 := \frac{N}{2}\left[\sqrt{{\mathbf p}_1^2+m^2} + \sqrt{{\mathbf p}_2^2+m^2} - (N-1)\frac{\kappa}{r_{12}}\right].
\label{Eq:H2}
\end{equation}
This expectation equality arises because the necessary boson
permutation symmetry of the the exact $N$-body ground state $\Psi$
implies that the expectations of the $N$ kinetic-energy terms in
$H$ are the same; and similarly for the $\half N(N-1)$
pair-potential terms. For convenience we have collected these into
$N/2$ times a two-body Hamiltonian; we have included the overall
$N/2$ factor and written this scaled two-body Hamiltonian as
$H_2.$ Thus with respect to the exact wavefunction $\Psi$ we may
write $E = \langle H\rangle = \langle H_2\rangle$, where $E$ is
the corresponding exact energy. If we denote by $E_2$ the bottom
of the spectrum of the two-boson problem with Hamiltonian $H_2$,
then, since boson symmetry in only two particles is in general a
weaker constraint than symmetry in all $N$ particles, it follows
that $E \ge E_2.$ Indeed, the dependence of $\Psi$ on the
variables $\{{\mathbf x}_3\dots {\mathbf x}_N\}$ that are not
present in $H_2$ cannot cause $\langle H_2\rangle$ to fall bellow
the bottom of the spectrum of $H_2.$ Thus $E_2$ provides a lower
energy bound to $E$. We shall sometimes express this as the
operator inequality $H \ge H_2$. A corresponding upper bound $E_g$
may be found with the aid say of a normalized Gaussian trial
function $\Phi_g$: thus $E \leq E_g = (\Phi_g, H\Phi_g).$ These
energy bounds allow us to compute bounds on the critical mass
$M_c,$ the largest allowed mass for such a bound system. In order
to make this point clear and to fix ideas we shall now compute an
explicit energy lower bound and from this a lower estimate to
$M_c.$ We first have to solve the two-body problem represented by
$H_2.$ If we consider for this problem new coordinates ${\mathbf
R} = {\mathbf x}_1 + {\mathbf x}_2,$ and ${\mathbf r} = {\mathbf
x}_1 - {\mathbf x}_2,$ then the corresponding momenta are related
by ${\mathbf p}_1 = {\mathbf p} + {\mathbf P}$ and ${\mathbf p}_2
= {\mathbf p} - {\mathbf P}$. If we introduce a vector ${\mathbf
k}$ which is orthogonal to ${\mathbf p}$ and ${\mathbf P}$, then
we may consider the following application of the triangle
inequality:
\begin{eqnarray*}
2(p^2 + m^2)^{\half} &= |2{\mathbf p} + 2m{\mathbf k}| \\
 & = |{\mathbf p}+{\mathbf P} + m{\mathbf k}+ {\mathbf p}-{\mathbf P} + m{\mathbf k}| \\
& \leq |{\mathbf p}_1 +m{\mathbf k}| + |{\mathbf p}_2 +m{\mathbf k}|.
\label{Eq:KE1}
\end{eqnarray*}
From this inequality and (\ref{Eq:H2}) we conclude the following
inequalities
\begin{equation*}
H \geq H_2 \geq N\left[\sqrt{p^2 + m^2} - \frac{(N-1)\kappa}{2
r}\right].
\end{equation*}
Consequently from (\ref{Eq:HLB}) we have
\begin{equation*}
E \geq Nm\left[1 -
\left(\frac{(N-1)\kappa\pi}{4}\right)^{2}\right] > Nm\left[1 -
\left(\frac{N\kappa\pi}{4}\right)^{2}\right].
\end{equation*}
We have replaced $N-1$ by $N$ merely for analytical convenience.
Thus we obtain the following $N$-boson lower energy bound
\begin{equation}
E \geq \frac{4m}{\pi\kappa}t\left(1-t^2\right)^{\half}, \quad t := \frac{N\kappa\pi}{4} \leq 1.
\label{Eq:LB1}
\end{equation}
It turns out that if we maximize the right-hand side of
(\ref{Eq:LB1}) with respect to $N$, that is to say, with respect
to the parameter $t,$ the critical value of $t$ is $\hat{t} =
1/\sqrt{2},$ so that the Herbst coupling inequality is satisfied
at the optimal point. Since mass and energy are identified in our
units, and $\kappa = Gm^2,$ we arrive at the bound $M_c >
(2/\pi)/Gm \approx 0.63662/Gm.$ This detailed calculation shows
how an energy bound leads to an estimate for the critical mass
$M_c.$ The principal goal of the article is to refine such
estimates. Thus we have here an explicit example of the phenomenon
under discussion: if $m$ is the mass of an alpha particle, say,
then $M$ cannot be larger than the mass of a modest mountain (we
shall present an upper bound to $M_c$ shortly). No such
possibility arises from the corresponding non-relativistic theory.

\section{Reduction: `equivalent' two-body problems}

The question of the implications of the necessary permutation
symmetry of the states for systems composed of many identical
particles is almost as old as quantum mechanics. There are a
number of historical threads. Before going into the technical
details of our problem, we shall briefly mention two of these. The
reasoning leading to the two-particle Hamiltonian $H_2$ suggests
that the energy depends on a reduced density matrix $\rho({\mathbf
x}_1,{\mathbf x}_2,{\mathbf x}'_1,{\mathbf x}'_2)$ obtained by
integrating $$\Psi({\mathbf x}_1,{\mathbf x}_2,{\mathbf
x}_3,\dots,{\mathbf x}_N)\Psi({\mathbf x}'_1,{\mathbf
x}'_2,{\mathbf x}_3,\dots,{\mathbf x}_N)$$over all the variables
${\mathbf x}_i$ with $i > 2.$ A question raised by this is, what
are the necessary features of $\rho$ which characterize it as
having come from an $N$-boson function $\Psi$? This is called the
$N$-representability problem and goes back at least to the early
papers of P.-O. L\"owdin \cite{Lowdin} and A. J. Coleman
\cite{Coleman}: a summary of early work in this direction can be
found in the introductory chapters of Ref.\cite{ACP} by A. J.
Coleman and M. Rosina. Much of the early work was concerned with
atomic and chemical systems. Density-matrix many-body theory and
the $N$-representability problem are still active areas of
research \cite{GidofalviM, ShenviW, LiuCV}. We note that, as in
the previous paragraph, one can derive energy lower bounds without
attempting to solve the $N$-representability problem
 generally.

Another story concerns nuclear type systems, where all the
particles enter the motion on an equal footing, and considerations
of centre-of-mass motion become important. In order to make this
point more explicit, the example of the harmonic oscillator is
helpful. We consider briefly the non-relativistic Hamiltonian
given by
\begin{equation}
H_{\rm HO} = \sum_{i=0}^{N} \frac{{\mathbf p}_i^2}{2m} + \sum_{1
=i <j}^N v r_{ij}^2. \label{Eq:NRHO}\end{equation} The earliest
treatment we know of for this problem is by Houston \cite{Houston}
in 1935; a solution expressed more specifically useful for our
purposes was found in 1953 by Post \cite{Post53}; the solubility
of the $N$-body harmonic-oscillator problem is periodically
rediscovered, with justifiable fresh enthusiasm. In units with
$\hbar = 1$ the bottom of the spectrum is given exactly by the
expression $E_{\rm HO} = 3(N-1)\sqrt{Nv/(2m)}.$ The exact
ground-state wave function is a Gaussian in the $N-1$ orthogonal
relative coordinates. If the same reasoning we used to derive the
semirelativistic operator bound $H \ge H_2$ above is now applied
to $H_{\rm HO},$ the resulting lower energy bound obtained is
exactly given by $E_L = E_{\rm HO}/\sqrt{2}.$ If, instead, the
`reduction' (to a two-body problem) is effected with Jacobi
relative coordinates, one obtains a lower bound for the harmonic
oscillator equal to $E_{\rm HO}$ itself. This type of reduction
has its own history. In is only possible to indicate a few key
events of this story in the present short article. In 1933, just
after the discovery of the neutron, physicists began to look at
few-nucleon problems. An approach emerged called the `equivalent
two-body method'. It was initiated by Wigner \cite{Wigner} and
employed by many researchers \cite{Feenberg, FeenbergK,FeenbergS,
MasseyB, BetheB,RaritaP} and eventually found its way into the
pages of Rosenfeld's book {\it Nuclear Forces} \cite{Rosenfeld} in
1948. The idea was always the same, to replace the $N$-body
problem by a tractable two-body problem. In many instances the
result yielded an energy lower bound, but this was unknown to the
workers at the time. The first rigorous results for such problems
came in 1956 when Post \cite{Post56} used Jacobi relative
coordinates to show that indeed a lower bound could be
constructed. In 1962 Post \cite{Post62} applied this bound to the
gravitational problem with pair potentials of the form
$-v/r_{ij}$; together with a Gaussian trial function, the energy
was determined to $18\%.$ Rigorous energy bounds with the aid of
Jacobi coordinates, and a discussion of the `equivalent two-body
method' may be found in a paper by Hall and Post \cite{Hall67a} in
1967. Some similar lower-bound results were later obtained by
Levy-Leblond \cite{Levy} and Stenschke \cite{Stenschke}. An
independent review and a certain sharpening of results by R. N.
Hill may be found in Ref.\cite{Hill}. The two streams of activity
intersected in a paper by Hall \cite{Hall67b} who used a similar
argument to that of Coleman \cite{Coleman} to obtain a lower bound
for a fermion system in the centre-of-mass frame in terms of a sum
over $N-1$ reduced two-particle energies; the bound was also
optimized over a certain set of allowed non-orthogonal relative
coordinates. Later this lower-bound theory (with non-orthogonal
relative coordinates) was extended to the excited states
\cite{Hall79}. A variety of alternative lower-bound models and
approaches have been developed, for example by Carr \cite{Carr78},
Manning \cite{Mann78}, and Balbutsev \cite{Bal81}. The
non-relativistic lower bound for the ground-state energy is
rediscovered from time to time, for example by Membrado {\it et
al.} \cite{MembradoPS} and Basdevant {\it et al.}
\cite{BasdevantMR}.

\section{Semirelativistic gravitating bosons}

We must now return to our main problem, the application of these
ideas to a semirelativistic many-body system. The complication
that the permutation symmetry (in its spatial aspect) is expressed
in the individual-particle coordinates whereas the wave function
is expressed in relative coordinates remains, and is adjoined by a
new difficulty, namely the non-locality of the semirelativistic
kinetic-energy operator. Here the $N$-body harmonic-oscillator
problem is now no longer solved exactly by the lower bound, but
with a finite error is less than $0.15\%$ \cite{HLS04}.

 The Hamiltonian $H,$ among others, has been adopted to
investigate spherically symmetric and nonrotating configurations
of purely gravitationally interacting bosons forming compact
objects known as ``boson
stars''\cite{M88,BMR89,MR89,Jetzer92,Raynal94}. This operator $H$
is composed of the relativistically correct expression for the
kinetic energy of all the involved bosons and {\em static\/}
potentials $\kappa/r_{ij}$ which describe the gravitational forces
between these particles. Therefore, it is clearly not possible in
this model to take into account retardation effects. In addition,
it goes without saying that this approach also omits
general-relativistic effects \cite{RuffiniB, FriedbergLP, BizonW}.
Sufficient conditions have been found both for relativistic
stability, which is characterized by the existence of a lower
bound on the Hamiltonian $H$ of (\ref{Eq:HN}), and for
relativistic gravitational collapse, which is inevitable if $H$ is
not bounded below. Moreover, semirelativistic bounds have been
derived for the maximum possible, or critical, mass $M_\mathrm{c}$
of boson stars, that is, the mass beyond which there must be
relativistic collapse.

The results of particular interest for this analysis can be
summarized as follows. The relativistic kinetic energy
$\sqrt{{\mathbf p}^2+m^2}$ satisfies\cite{M88} a
(tangential\cite{HLS01}) operator inequality, involving an
arbitrary real parameter $\mu$ with the dimension of
mass:\[\sqrt{{\mathbf p}^2+m^2}\le\frac{{\mathbf
p}^2+m^2+\mu^2}{2\mu}\quad\forall\ \mu>0.\]This inequality can be
adopted to relate the semirelativistic Hamiltonian
$H,$~Eq.~(\ref{Eq:HN}), to its nonrelativistic counterpart. A
variational bound on the ground-state energy of the
nonrelativistic $N$-particle problem therefore translates into the
upper bound $M_\mathrm{c}<1.52/Gm$\cite{BMR89}. Exploiting the
(only numerically computed) nonrelativistic ground-state energy,
this bound is refined to $M_\mathrm{c}<1.518/Gm$\cite{Jetzer92}.
Rewriting $H$ as a sum of one-particle Hamiltonians, each of which
is bounded from below~by the lowest positive eigenvalue of the
Klein--Gordon Schr\"odinger equation with Coulomb potential,
yields a bound\footnote{Ref.\cite{HL05} discusses this simple
``$N/2$ lower energy bound'' for arbitrary potentials.} to the
bottom $E$ of the spectrum of~$H$\cite{MR89}:\begin{equation}E\ge
Nm\sqrt{\frac{1+\displaystyle\sqrt{1-(N-1)^2\kappa^2}}{2}}
,\quad(N-1)\kappa<1.\label{Eq:MRSLB}\end{equation}The replacement
of $N-1$ under the square root by $N$ slightly weakens this bound
but allows for its analytic maximization, which entails the
analytic lower bound\cite{MR89}
\[M_\mathrm{c}\ge\frac{4}{3\sqrt{3}Gm}\simeq\frac{0.7698}{Gm}.\]
Together these estimates constrain $M_\mathrm{c}$ to the range
$0.7698<GmM_\mathrm{c}<1.518.$ The resulting ratio of upper to
lower bounds on $M_\mathrm{c}$ is $r_\mathrm{U/L}\simeq2.0.$ The
so-called local-energy theorem may be used to increase the lower
bound, whereas a more sophisticated choice of trial functions
diminishes the variational upper~bound. The combined effect of
these improvements is to narrow down the range for~$M_\mathrm{c}$
to $0.8468<GmM_\mathrm{c}<1.439,$ with upper- to lower-bound ratio
of $r_\mathrm{U/L}\simeq1.7$\cite{Raynal94}.

We have re-analysed the upper bound of
Ref.\cite{Raynal94} with positive non-monotone Hartree
wave-function factors $\phi(r).$ With the factor (before scale
optimization) $\phi(r)=ce^{-r}(1+ar),$ $a>0,$ we confirm the
findings\cite{Raynal94} that the best value of $a$ is about
$a\simeq1,$ which yields $GmM_\mathrm{c}<1.43871\approx1.439.$
This $\phi$ does indeed seem to be close to the best possible
Hartree factor. With $a=1.13,$ we get a slight improvement, viz.,
$GmM_\mathrm{c}<1.43854.$ With the factor
$\phi(r)=ce^{-r}(1-be^{-r})$~we obtain our best result, viz.,
$GmM_\mathrm{c}<1.43764$ for $b=0.625.$ Thus we have been able to
lower the upper bound on the critical mass slightly to
$M_\mathrm{c}<1.438/Gm.$

In this paper we tighten the interval allowed for the critical
mass of boson~stars, by employing an improved {\em analytic\/}
lower bound\cite{HL05} on the ground-state~energy of the
$N$-particle Hamiltonian (\ref{Eq:HN}) for semirelativistic
self-gravitating $N$-boson systems, to a range characterized by an
upper- to lower-bound ratio $r_\mathrm{U/L}\simeq1.3.$ The region
of validity of a lower bound on the Hamiltonian $H$ defines
the~range of relativistic stability of the gravitating
$N$-particle system under study\cite{MR89}:~our improved lower
energy bound discussed below increases somewhat the stability
region obtained in Ref.\cite{MR89}; for instance, for couplings
$\kappa\ll1$ --- which allows~for large $N$ --- this increase of
the stability range amounts to an 11\% improvement.

\section{Lower bound for self-gravitating semirelativistic
$N$-boson systems}\label{}Let $|\Psi\rangle,$
$\langle\Psi|\Psi\rangle=1,$ represent the normalized ground state
of $H,$ corresponding to its lowest eigenvalue
$E\equiv\langle\Psi|H|\Psi\rangle.$ Now, the bosonic nature of the
identical bound-state constituents forces the eigenstates of $H$
(i.e., their wave functions) to be totally symmetric under any
permutation of the individual-particle coordinates $\{{\bf
x}_1,{\bf x}_2,\dots,{\bf x}_N\}.$ The boson permutation symmetry
of $|\Psi\rangle$ reduces the $N$-body problem posed by the
Hamiltonian $H$ to a {\em constrained\/} two-particle
problem\cite{HLS04}:
\begin{equation}E=\left\langle\Psi\left|N\sqrt{{\mathbf
p}_N^2+m^2}-\frac{\gamma\kappa}{r_{N-1,N}}\right|\Psi\right\rangle,
\quad\gamma\equiv\frac{N(N-1)}{2}.\label{Eq:EBPS}\end{equation}

By use of permutation symmetry, Eq.~(\ref{Eq:EBPS}) may be cast
into the equivalent form\begin{equation}
E=\left\langle\Psi\left|\frac{N}{2}\left(\sqrt{{\mathbf
p}_1^2+m^2}+\sqrt{{\mathbf p}_2^2+m^2}\right)
-\frac{\gamma\kappa}{r_{12}}\right|\Psi\right\rangle.\label{Eq:EBPS1}
\end{equation}After removal of the center-of-mass momentum from
${\mathbf p}_1$ and ${\mathbf p}_2,$ this apparent two-particle
problem reduces to a one-body problem in the relative coordinate
and momentum of the particles 1, 2 for which the Klein--Gordon
equation with Coulomb interaction gives a lower bound: this
eventually yields the bound (\ref{Eq:MRSLB}).

The lower bound (\ref{Eq:MRSLB}), however, is dramatically
improved\cite{HL05} by the use of Jacobi relative coordinates. The
transformation from a given set $\{{\bf x}_i,\ i=1,2,\dots,N\}$ of
coordinates to another set $\{\mbox{\boldmath{$\rho$}}_k,\
k=1,2,\dots,N\}$ may be defined by a matrix $B=(B_{ki})$:
$\mbox{\boldmath{$\rho$}}=B{\mathbf x}.$ The orthogonality
$B^{-1}=B^{\mathrm T}$ of $B$ is not mandatory but may prove to be
convenient. The momenta $\{\mbox{\boldmath{$\pi$}}_i\}$ conjugate
to the $\{\mbox{\boldmath{$\rho$}}_i\}$ are~then also determined
by $\mbox{\boldmath{$\pi$}}=(B^{-1})^{\rm T}{\mathbf p}=B{\mathbf
p}.$ The transformation to Jacobi relative coordinates is
represented by an orthogonal matrix with the first row given
by\[B_{1i}=\frac{1}{\sqrt{N}}\quad\forall\
i=1,2,\dots,N,\]whereas, for all $2\le k\le N,$ in the $k$th row
only the first $k$ entries are nonzero:
\begin{eqnarray*}B_{ki}&=&\frac{1}{\sqrt{k(k-1)}}\quad\forall\
i=1,2,\dots,k-1,\\B_{kk}&=&-\sqrt{\frac{k-1}{k}},\\B_{ki}&=&0\quad
\forall\ i=k+1,k+2,\dots,N.\end{eqnarray*}Evidently, by the
definition of $B$ its first row generates the usual center-of-mass
variable $\mbox{\boldmath{$\rho$}}_1,$ while its second row
introduces a pair distance $\mbox{\boldmath{$\rho$}}_2=({\mathbf
x}_1-{\mathbf x}_2)/\sqrt{2}.$

Any boson state $|\Phi\rangle$ is symmetric under permutations of
all {\em individual-particle\/} coordinates. However, a
non-Gaussian boson state is not necessarily symmetric in the
Jacobi {\em relative\/} coordinates. Nevertheless, as has been
shown in App.~A of Ref.\cite{HL05}, each such $|\Phi\rangle$
satisfies, for all $i,k\ge2,$ the $N$-representability~identities
\begin{eqnarray}
\langle\Phi|\mbox{\boldmath{$\rho$}}_i\cdot\mbox{\boldmath{$\rho$}}_k|\Phi\rangle&=&
\delta_{ik}\left\langle\Phi\left|\mbox{\boldmath{$\rho$}}_2^2\right|\Phi\right\rangle,
\nonumber\\
\langle\Phi|\mbox{\boldmath{$\pi$}}_i\cdot\mbox{\boldmath{$\pi$}}_k|\Phi\rangle&=&
\delta_{ik}\left\langle\Phi\left|\mbox{\boldmath{$\pi$}}_2^2\right|\Phi\right\rangle.
\label{Eq:Nrepid}\end{eqnarray}

Now, for the sake of notational simplicity, let us introduce some
abbreviations:\begin{eqnarray*}&&\lambda\equiv\frac{N-1}{N},\quad
a\equiv\frac{1}{\sqrt{\lambda}}=\sqrt{\frac{N}{N-1}},\\
&&b\equiv\sqrt{\frac{N-2}{N-1}},\quad
c\equiv\frac{b}{a}=\sqrt{\frac{N-2}{N}}.\end{eqnarray*}These
parameters $\lambda,a,b,c$ are, of course, related by $a^2+b^2=2$
and $1+c^2=2\lambda.$ In terms of Jacobi relative coordinates, the
expectation value (\ref{Eq:EBPS}) then becomes
\[E=\left\langle\Psi\left|N\sqrt{\left(a\mbox{\boldmath{$\pi$}}_1
-\sqrt{\lambda}\mbox{\boldmath{$\pi$}}_N\right)^2+m^2}-
\frac{\gamma\kappa}{|a\mbox{\boldmath{$\rho$}}_N-b\mbox{\boldmath{$\rho$}}_{N-1}|}
\right|\Psi\right\rangle.\]We assume that the eigenstate
$|\Psi\rangle$ depends on $\{\mbox{\boldmath{$\rho$}}_2,
\mbox{\boldmath{$\rho$}}_3,\dots,\mbox{\boldmath{$\rho$}}_N\}$ but
not on $\mbox{\boldmath{$\rho$}}_1.$ A lemma shown in
Ref.\cite{HLS02} allows us to remove the center-of-mass momentum
$\mbox{\boldmath{$\pi$}}_1$ from the kinetic term. Thus the
$N$-body ground-state energy $E$ simplifies~to\begin{equation}
E=\left\langle\Psi\left|N\sqrt{\lambda\mbox{\boldmath{$\pi$}}_N^2+m^2}-
\frac{\gamma\kappa}{|a\mbox{\boldmath{$\rho$}}_N-b\mbox{\boldmath{$\rho$}}_{N-1}|}
\right|\Psi\right\rangle.\label{Eq:E(N,N-1)}\end{equation}

Focusing to the $(N-1,N)$ subsystem we introduce new coordinates
$\{{\mathbf R},{\mathbf r}\}$ and their conjugate momenta
$\{{\mathbf P},{\mathbf p}\},$ by performing the coordinate
transformation\[\left(\begin{array}{c}{\mathbf R}\\{\mathbf
r}\end{array}\right)=O\left(\begin{array}{l}\mbox{\boldmath{$\rho$}}_N\\
\mbox{\boldmath{$\rho$}}_{N-1}\end{array}\right),\quad
\left(\begin{array}{c}{\mathbf P}\\{\mathbf
p}\end{array}\right)=\frac{O}{2}\left(\begin{array}{l}\mbox{\boldmath{$\pi$}}_N\\
\mbox{\boldmath{$\pi$}}_{N-1}\end{array}\right).\]The expectation
value (\ref{Eq:E(N,N-1)}) suggests the most favourable choice of
the matrix~$O$:
\[O\equiv\left(\begin{array}{rr}b&a\\a&-b\end{array}\right)=O^{\rm
T},\quad O^{\rm T}O=O^2=2.\]Upon this change of variables, the
ground-state energy $E$ of the Hamiltonian $H$ is given by the
expectation value $E=\langle\Psi|{\mathcal H}|\Psi\rangle$ of the
two-particle Hamiltonian\[{\mathcal H}\equiv N\sqrt{({\mathbf
p}+c{\mathbf P})^2+m^2}- \frac{\gamma\kappa}{r},\quad
r\equiv|{\mathbf r}|.\]It may be proved that ${\mathcal H}$ is
bounded from below by the Hamiltonian entering~in the expectation
value on the right-hand side of Eq.~(\ref{Eq:EBPS1}) (see App.~B
of Ref.\cite{HL05}).

For the new momenta ${\mathbf P}$ and ${\mathbf p},$ the
identities (\ref{Eq:Nrepid}) translate into the constraints\[
\left\langle\Phi\left|{\mathbf P}^2\right|\Phi\right\rangle=
\left\langle\Phi\left|{\mathbf p}^2\right|\Phi\right\rangle
\quad\mbox{and}\quad\langle\Phi|{\mathbf P}\cdot{\mathbf
p}|\Phi\rangle=0.\]Consequently, we have to look for the bottom of
the spectrum of the constrained problem posed by the operator
${\mathcal H}$ in a domain ${\mathcal D}$ restricted by these
conditions:\[{\mathcal D}=\left\{|\varphi\rangle\in
L^2(\Re^6): \left\langle\varphi\left|{\mathbf
P}^2\right|\varphi\right\rangle= \left\langle\varphi\left|{\mathbf
p}^2\right|\varphi\right\rangle,\ \langle\varphi|{\mathbf
P}\cdot{\mathbf p}|\varphi\rangle=0\right\}.\]This bottom
${\mathcal E}$ of the spectrum of ${\mathcal H},$ of course,
provides a lower bound to $E$:

\[E=\langle\Psi|{\mathcal H}|\Psi\rangle
\ge\inf_{{|\varphi\rangle\in{\mathcal D}}\atop{
\langle\varphi|\varphi\rangle=1}}\langle\varphi|{\mathcal
H}|\varphi\rangle\equiv{\mathcal E}.\]

Let $|\psi\rangle\in{\mathcal D},$ $\langle\psi|\psi\rangle=1,$ be
the eigenstate of the Hamiltonian ${\mathcal H}$ corresponding to
this lowest eigenvalue ${\mathcal E}.$ The eigenvalue equation of
${\mathcal H}$ satisfied by $|\psi\rangle$ reads
\[N\sqrt{({\mathbf
p}+c{\mathbf P})^2+m^2}\,|\psi\rangle=\left({\mathcal
E}+\frac{\gamma\kappa}{r}\right)|\psi\rangle.\]By squaring this
relation and remembering the constraints that define ${\mathcal
D}$ we~get\begin{equation}{\mathcal
E}^2-N^2m^2=4\gamma\left\langle\psi\left|{\mathbf p}^2
-\frac{\kappa{\mathcal
E}}{2r}-\frac{\gamma\kappa^2}{4r^2}\right|\psi\right\rangle.
\label{Eq:Kratzer}\end{equation}Now, by assumption, $|\psi\rangle$
is the lowest eigenstate of ${\mathcal H}$ but not
necessarily~of~the one-particle Kratzer-type\cite{Kratzer}
operator in Eq.~(\ref{Eq:Kratzer}). According to the variational
principle, the (well-known) lowest eigenvalue of this Kratzer-type
Hamiltonian provides a lower bound on the expectation value in
Eq.~(\ref{Eq:Kratzer}). Solving the implicit inequality for
${\mathcal E}$ yields a lower bound to ${\mathcal E},$ and thus to
$E$\cite{HL05}; this lower bound is nothing but the lowest
positive eigenvalue of the corresponding Klein--Gordon
Schr\"odinger equation\cite{Schroedinger} for gravitational
interaction of appropriate~strength:\begin{equation}E\ge
Nm\sqrt{\frac{1+\displaystyle\sqrt{1-\gamma\kappa^2}}{2}},\quad
\gamma\kappa^2<1.\label{Eq:HLILB}\end{equation}Our improved lower
bound (\ref{Eq:HLILB}) on the ground-state energy of any
self-gravitating $N$-boson system is of the same form as the bound
(\ref{Eq:MRSLB}) but with $\gamma\equiv N(N-1)/2$ replacing
$(N-1)^2,$ which is favourable since $N(N-1)/2<(N-1)^2$ for $N>2.$

\section{Semirelativistic stability and critical mass of boson
stars}Let us now analyze the implications of the improved lower
energy bound~(\ref{Eq:HLILB})~for both stability against
gravitational collapse and maximum mass of boson stars.

The existence of a lower bound on the spectrum of the Hamiltonian
operator~$H$ guarantees the stability of the self-gravitating
boson system against relativistic gravitational collapse. The
region of validity of such kind of lower energy bound delimits the
stability range of the bound state described by $H.$ By
construction, our bound (\ref{Eq:HLILB}) holds for all $N$
satisfying $N(N-1)\kappa^2<2.$ This~stability~region is larger
than the one, $(N-1)\kappa<4/\pi,$ found in Ref.\cite{MR89}. For
large values of~$N,$ allowed for sufficiently small couplings
$\kappa,$ this gain amounts to $\pi/2\sqrt{2}=1.11.$ In terms of
Newton's constant $G$ and the particle mass $m,$ a {\em sufficient
condition for relativistic stability\/} thus is that the particle
number $N$ fulfils the constraint\[N(N-1)<\frac{2}{(Gm^2)^2}.\]

Following Ref.\cite{MR89}, in order to allow for a discussion by
elementary methods,~we weaken Eq.~(\ref{Eq:HLILB}) by replacing
the exact $N$ dependence $\gamma\equiv N(N-1)/2$ by $N^2/2$:\[E\ge
Nm\sqrt{\frac{1+\displaystyle\sqrt{1-N^2\kappa^2/2}}{2}},\quad
N\kappa<\sqrt{2}.\]Evidently, even this weakened lower bound is
still above the lower bound~(\ref{Eq:MRSLB})~for all
$N>2+\sqrt{2}\simeq3.41;$ for large $N,$ the weaker bound
approaches the exact~one. The (single) maximum of this weakened
lower bound is situated at the critical point $\hat N=4/3\kappa,$
which is, fortunately, in the interior of the region of validity
of our lower bound (\ref{Eq:HLILB}) on the Hamiltonian $H$ as
$\hat N<\sqrt{2}/\kappa.$ This maximum~thus constitutes the
(improved) lower bound on the critical mass $M_\mathrm{c}$ of
boson stars
\begin{equation}M_\mathrm{c}\ge\frac{4\sqrt{2}m}{3\sqrt{3}\kappa}
=\frac{4\sqrt{2}}{3\sqrt{3}Gm}=\frac{1.08866}{Gm}.\label{Eq:Mc}
\end{equation}This lower $M_\mathrm{c}$ bound is larger by exactly
a factor $\sqrt{2}$ than the result of~Ref.\cite{MR89}. Combining
the lower bound (\ref{Eq:Mc}) with the Rayleigh--Ritz upper bound
of Ref.\cite{Raynal94} tightens the (Newtonian-limit) prediction
for $M_\mathrm{c}$ to $1.08866<GmM_\mathrm{c}<1.439,$ reducing
thus the ratio between upper and lower bounds on $M_\mathrm{c}$ to
$r_\mathrm{U/L}\simeq1.3.$

In summary, with the aid of an improved lower bound\cite{HL05}
(based on the~relative coordinates of the bound-state
constituents) on the bottom of the spectrum of the
semirelativistic $N$-boson Hamiltonian (\ref{Eq:HN}) with
gravitational interaction we have succeeded in enlarging the range
of semirelativistic stability of boson~stars and in halving the
theoretical uncertainty in the maximum mass of boson stars.

\section*{Acknowledgements}

One of us (RLH) gratefully acknowledges both partial financial support
of this research under Grant No.\ GP3438 from the Natural Sciences
and Engineering Research Council of Canada and the hospitality of
the Institute for High Energy Physics of the Austrian Academy of
Sciences, Vienna, where part of the work was done.

\hfil\vfil\break

\end{document}